# Multi-directional cloak design by all dielectric unit-cell optimized structure

Muratcan Ayik[1,2], Hamza Kurt[3], Oleg V. Minin[4], Igor V. Minin[4] and Mirbek Turduev[5,*]

[1]*Department of Electrical and Electronics Engineering, Middle East Technical University, Ankara 06800, Turkey*
[2]*Aselsan Inc., Ankara 06200, Turkey*
[3]*School of Electrical Engineering, Korea Advanced Institute of Science and Technology (KAIST), Daejeon, 34141, Republic of Korea*
[4]*Nondestructive school, Tomsk Polytechnic University, 634050, Tomsk, Russia*
[5]*Department of Electrical and Electronics Engineering, Kyrgyz-Turkish Manas University, Bishkek 720038, Kyrgyzstan*
*\*Corresponding author: mirbek.turduev@manas.edu.kg*

**Abstract**

In this manuscript, we demonstrate the design and experimental proof of an optical cloaking structure which multi-directionally conceals a perfectly electric conductor (PEC) object from an incident plane wave. The dielectric modulation around the highly reflective scattering PEC object is determined by an optimization process for multi-directional cloaking purposes. And to obtain the multi-directional effect of the cloaking structure, an optimized slice is mirror symmetrized through a radial perimeter. Three-dimensional (3D) finite-difference time-domain method is integrated with genetic optimization to achieve cloaking design. In order to overcome the technological problems of the corresponding devices in the optical range and to experimentally demonstrate the proposed concept, our experiments were carried out on a scale model in the microwave range. The scaled proof-of-concept of proposed structure is fabricated by 3D printing of polylactide material, and the brass metallic alloy is used as a perfect electrical conductor for microwave experiments. A good agreement between numerical and experimental results is achieved. The proposed design approach is not restricted only to multi-directional optical cloaking but can also be applied for different cloaking scenarios dealing with electromagnetic waves in nanoscales as well as other types of such as acoustic waves. Using nanotechnology, our scale proof-of-concept research will take the next step towards the creation of "optical cloaking" devices.

## 1. Introduction

One of the most intriguing and mysterious phenomena that have captured researchers' imaginations for centuries is invisibility. In general, invisibility means that light wave incident on the object should remain its optical property after passing through an object. In other words, the scattered and deteriorated field resulting from the object should be reconstructed and corrected to replicate the incident wave. The first realistic idea of the invisibility cloak has been introduced to the literature by the pioneering works of Leonhardt and Pendry [1, 2]. Here, application of conformal mapping concept is proposed and the first application of transformation optics (TO) into optical cloaking is demonstrated [3-6]. In TO, the incident wave is guided to follow a curved trajectory by simply bending the coordinate system to obtain the cloaking of objects.

Along with TO, an interesting approach for concealing an object named as "carpet-cloaking" is proposed. Here large scatterer object is hidden under a reflective layer named carpet by using quasi-conformal mapping [7-11]. Similarly, as in TO, this non-Euclidian approach provides an interesting solution to prevent an object from detection [12]. This approach has been further developed to design and analyze carpet cloaking methods [13-15]. Lately, different from the TO concept, new design strategies for optical cloaking are proposed such as metasurfaces [16,17], zero-refractive-index materials [18], plasmonics [19-21], woodpile photonic structures [22], graded index structures [23], and superluminal media [24] to operate in different wavelength regimes including microwave, terahertz, infrared and visible. Also, the mantle cloaking technique offers scattering cancellation by covering the cylindrical dielectric with conducting helical sheet [25] or concentric mantle cloak [26] by satisfying surface impedance condition at designed frequencies.

Moreover, to obtain optical cloaking effect the suppression of scatterings resulting from an object is realized by using generalized Hilbert transforms [27] and Kramers-Kronig relations [28]. In addition to these studies, the focusing effect is also used for creating invisible regions both in ray and wave optics [29, 30]. Finally, the idea of using optimization algorithms for the generation/reshaping of the cloaking region shows promising results [31-34]. Here, the optimization methods search for possible designs of cloaking structures in accordance with a specific objective function. Furthermore, experimental verifications at microwave frequency regimes of cloaking designs based on optimization methods were reported in Refs. [35] and [36].

The nanotechnology plays a significant role in the development and creation of new cloaking devices in nanoscale [37]. Moreover, optical cloaking plays an important role in industry where the development of nanotechnology makes possible the design of novel camouflage systems and radar absorbing surfaces for low observable technologies [38 - 41]. In accordance with the state-of-the-art nanofabrication technology, controlling the flow of light along with their spatial mapping at the nanoscale in some cases is always not possible. On the other hand, thanks to the scalability of the Maxwell's equations [42, 43], one can always analyze the designed prototype at the microwave region for verification of the proof of the proposed concept [44 - 46].

In this study, we propose the design of all-dielectric, lossless, broadband, and passive multi-directional cloaking structure which conceals a high reflective perfectly electric conductor (PEC) material/object from an incident plane wave. The designed cloak is composed of polylactide (PLA) material which is a low loss biodegradable thermoplastic polymer with a low permittivity value. This dielectric material is widely used in three-dimensional (3D) additive printing technology and gives the opportunity for direct and cost-effective fabrication of the devices. The generalized framework of the proposed design approach with numerical and experimental analysis of the performance of the designed cloaking structure is provided in the current study. In addition, experimental verification of numerical results is performed at microwave frequency regime at around 10 GHz to demonstrate the operating principle of the design. As it was noted above the scale model of the proposed cloaking structure allow realizing a "rapid low-cost prototyping" for verification of proof-of-concept in microwave regime. Also, the physical mechanism of directional concealing effect of the designed optical cloak is primarily associated with the imperfect conformal mapping and partial suppression of scattered fields from the object. Since complete cloaking is impossible by conformal mapping with realistic material parameters, the remaining scattering is eliminated by an intelligent rendering of the cloaking structure thanks to advanced optimization. In addition, the proposed design methodology can find various cloaking applications of electromagnetic waves and may enable the multi-directional concealment of different objects possessing various sizes and shapes.

## 2. Design steps and numerical results

To optically hide object or to make it invisible, the incident wave should be reconstructed without distortion after passing through an object. In other words, the scattered field resulting from the object should be corrected/transformed to replicate the incident wave. For this reason, one should design such an environment around object that enables suppression of scatterings and diminishing of back-reflections. Hence, in contrast to forward design approaches the problem of optical cloaking can be treated as an inverse problem. Here the desired optical properties of the output electromagnetic field are defined and integrated into the cost function of the optimization method, and the algorithm iteratively searches for the best structure (environment) that provides the desired output. Therefore, to obtain desired optical cloaking effect, we optimally modulated the cloaking region's effective index distribution.

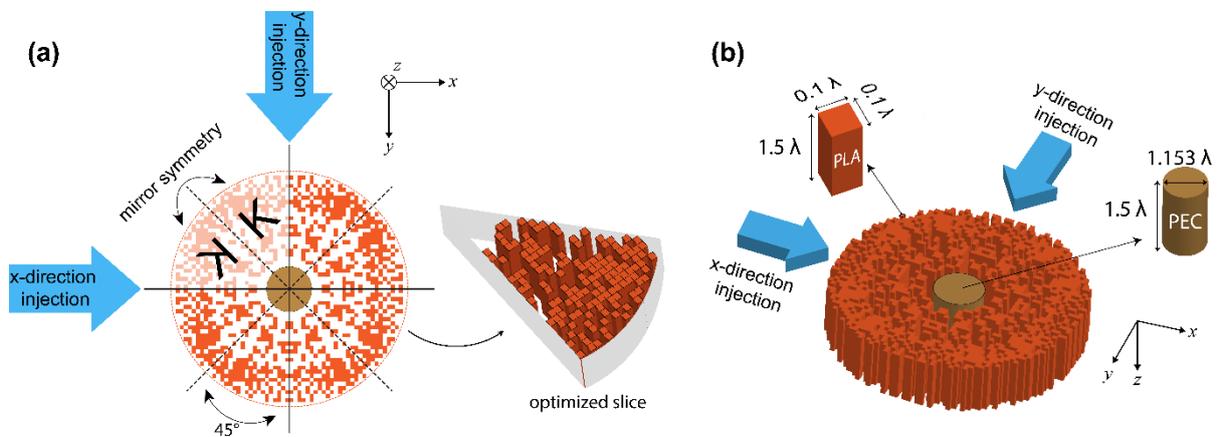

**Figure 1**. *(a) Schematic representation and the design approach of the cloaking structure and (b) three-dimensional view of the designed cloaking structure with physical dimensions of each unit cell and the PEC object. The letter "K" indicates applied symmetry effect to the structure.*

In this study, to achieve an optical cloaking effect, the concept of covering highly scattering material by index modulated structure is considered. For this reason, the circular shape is selected for the covering structure, which is also beneficial for multi-directional optical concealing in both *x*- and *y*- propagation directions. Fig. 1(a) illustrates the schematic representation of the design approach. As can be seen in Fig. 1(a), the circular region is divided into 8 slices with 45° internal angles to increase the directional independency. Here, to obtain bi-directional effect of the cloaking structure, optimized slice is mirror symmetrized through radial perimeter (the letter "*K*" shows the symmetry effect). In other words, by this symmetry concept the proposed structure demonstrates the exact same optical light conveying behavior in both injection *x*- and *y*- directions. The dashed and solid lines superimposed on the schematics in Fig. 1(a) define the border lines of mirror symmetry and rotational symmetry, respectively. In other words, the designed structure has 8 mirror symmetry slices which provide 4-fold rotational symmetry with 90° rotation angles. Here, rotational symmetry border lines define the injection directions of the cloak which are defined by blue arrows in Fig. 1(a). It should be noted that it is also possible to increase the number of injection directions (for multi-directional cloaking) by properly increasing the number of symmetry slices. Symmetry slices considered to be composed of rectangular shaped unit cells. The unit cells can be in two different states such as PLA ($\varepsilon_{PLA}$-existence of the unit cell) or Air ($\varepsilon_{air}$-absence of the unit cell) according to the decision of the applied optimization.

It is important to note that to define the states of those unit cells the genetic algorithm (GA) is integrated with the 3D finite difference time domain (FDTD) method [47]. GA is used for optimal distribution of permittivity that reduces observability of the object. GA is an evolutionary algorithm, i.e., a meta-heuristic, that mainly adapts advantage of the survival of the fittest in the evolutionary process. As the biological counterparts of evolution theory, GA comprises mechanisms such as crossover, mutation, and selection. GA iteratively searches the solution space to find candidate solutions to the problem described as a cost function. Here, GA decides whether each unit cell inside the optimization region is filled with PLA material or not. The algorithm fills the unit cell with air if it generates the binary number "0". Otherwise, for "1", it fills the unit cell with PLA. The three-dimensional view of the optimized cloaking structure with materials' parameters is demonstrated in Fig. 1(b). All unit cells filled with PLA are structurally identical and each one emerges as a rectangular prism that has dimensions of 0.1λ×0.1λ×1.5λ as shown as an inset in Fig. 1(b). Throughout the study the dielectric constants of PLA material and air are fixed to $\varepsilon_{PLA} = 2.4$, and $\varepsilon_{air} = 1.0$, respectively. In place of the highly scattered object which intended to be hidden from the incident wave, the cylindrical shaped perfect electrical conductor (PEC) is considered. The cylindrical PEC has a diameter of 1.15λ and a height of 1.5λ. Additionally, the diameter of the final structure is measured as 5.6λ. Considering the dimension of the structure we can say that designed structure belongs to the class of mesotronics [48].

The main goal of the study is to design such a surrounding structure that reduces the scattering effect of itself and the PEC that located inside the cloaked region. For this reason, before starting the optimization it is instructive to inspect the incident light scattering effect of bare PEC without cloaking, PEC coated with a fully filled/solid structure and PEC coated with a randomly PLA filled structure.

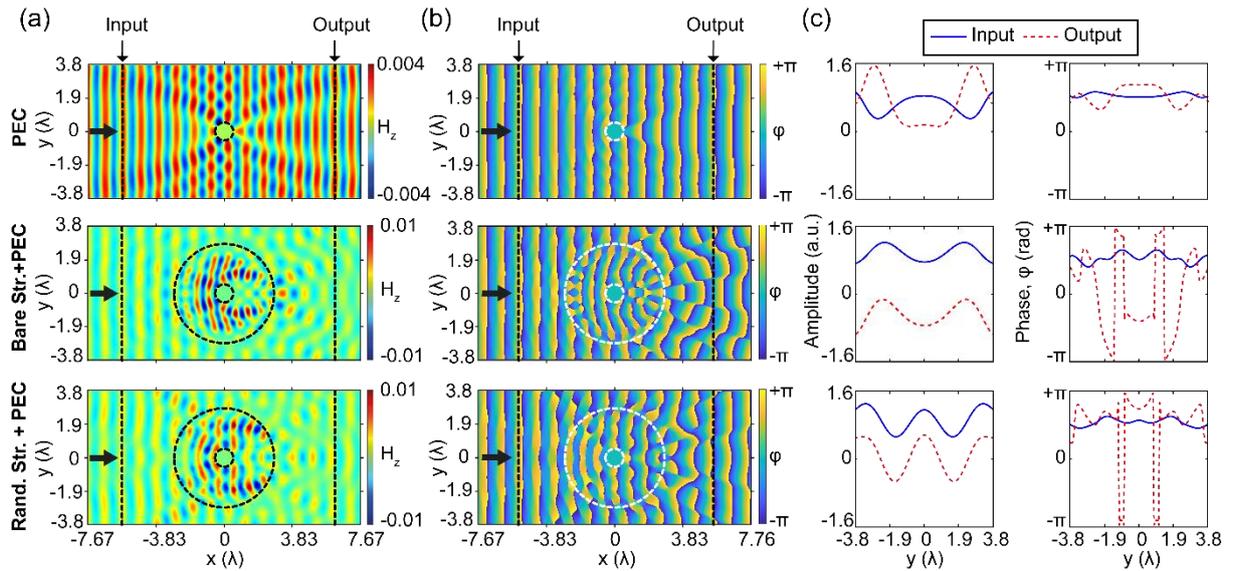

**Figure 2**. *The numerically calculated (a) magnetic field, (b) phase distributions, and (c) their cross-sectional amplitude and phase profiles at the front and back cross sections for the PEC, fully filled structure with PEC and a randomly filled structure with PEC, respectively from top to bottom. The black arrows indicate the incident waves which propagate in the x-direction. The dashed circles represent the*

*boundaries of the PEC material and the obtained structures. The position profiles at the input and output of the cross sections are signified by the vertical dashed lines. All calculations were performed at 10 GHz.*

For numerical analysis of light matter interaction, the 3D FDTD is employed and as an incident light source transverse electrical (TE) polarized plane wave is considered. It should be noted that, for TE polarization, the electric field components are along the *xy*-plane ($E_x$, $E_y$) and the magnetic field ($H_z$) is perpendicular to the *xy*-plane. The corresponding results are gathered in Fig. 2. The calculated magnetic field distributions as well as phase distributions for all structures are given in Figs. 2(a) and 2(b) (structure types of the corresponding fields are labeled on a vertical axis), respectively. Moreover, cross-sectional field and phase profiles are extracted before ("*input*") and after ("*output*") structure in propagation *x*-direction at the distance of 6.2λ from the center of the cloak (which corresponds to radiative near-field region if we consider cloak center as a wave source) [51]. The corresponding input and output cross-sectional profiles indicate the level of distortions in the wave fronts of the propagated light as seen in Fig. 2(c). As can be seen from the figure plots, while the bare PEC is a small object compared to the size of its coating structure, it strongly scatters the incident wave. The variations in the output cross-sectional profile are higher than the input side due to the scantiness of back reflections as seen in Fig. 2(c). On the other hand, without any optimization process when the cloaked object is coated with solid PLA material, it is obvious that the scattering effect of bare PEC worsens due to expanding light matter interactions into the reckless distribution of covering dielectric material. Also, in the case of randomly distributed cloaking structure, one can see that chaotically superimposed higher order modes are enhancing back reflections by distorting phase profiles at the input and output locations of the structure. All these undesired effects motivate us to use optimized index modulation of the PEC covering material to design a cloaking structure with negligible scattering characteristics.

The GA is employed to optimize the cloaking region by reducing the scattering of the transmitted and reflected fields. It minimizes the cost function given below:

$$f_{cost} = H_{error} + \varphi_{error} - T. \qquad (1)$$

In Eq. 1, $H_{error}$ and $\varphi_{error}$ stand for the total wave front distortions of magnetic field and phase field distributions at the input and output planes, respectively. Here we tried to conserve the linearity of the input and output crosses by minimizing the differences between the cross-sectional field and phase profiles and their average values. These differences were defined as distortion errors for both magnetic field distributions and phase profiles. The error for magnetic field profile is formulated as follows:

$$H_{error} = \sum_y (|H_z(x_{in}, y) - \overline{H_z}(x_{in}, y)| + |H_z(x_{out}, y) - \overline{H_z}(x_{out}, y)|), \qquad (2)$$

where, $H_z$ is the magnetic field and $\overline{H_z}$ represents the mean value of the magnetic field at cross section location of $x_{in}$ and $x_{out}$ which denote the input and output profiles' positions, respectively. Also, $y$ stands for the order of the mesh cells along the *y*-axis in the simulation area. The same concept is also used for the error calculation of phase profiles as follows:

$$\varphi_{error} = \sum_y (|\varphi_z(x_{in}, y) - \overline{\varphi_z}(x_{in}, y)| + |\varphi_z(x_{out}, y) - \overline{\varphi_z}(x_{out}, y)|), \qquad (3)$$

where, $\varphi_z$ represents cross-sectional phase profile and $\overline{\varphi_z}$ represents the mean value of the phase front at the predefined locations. In addition, in Eq. 1 the $T$ represents the transmission efficiency percentage which measures the total optical power of the incident wave is transmitted through the structures. Besides correction of distorted fields, the maximization of transmission efficiency is also important because the proper optical cloaking is only reasonable with high transmission efficiency. Here high transmissivity has an important impact on the transparency characteristic of the cloak which minimizes its shadowing effect.

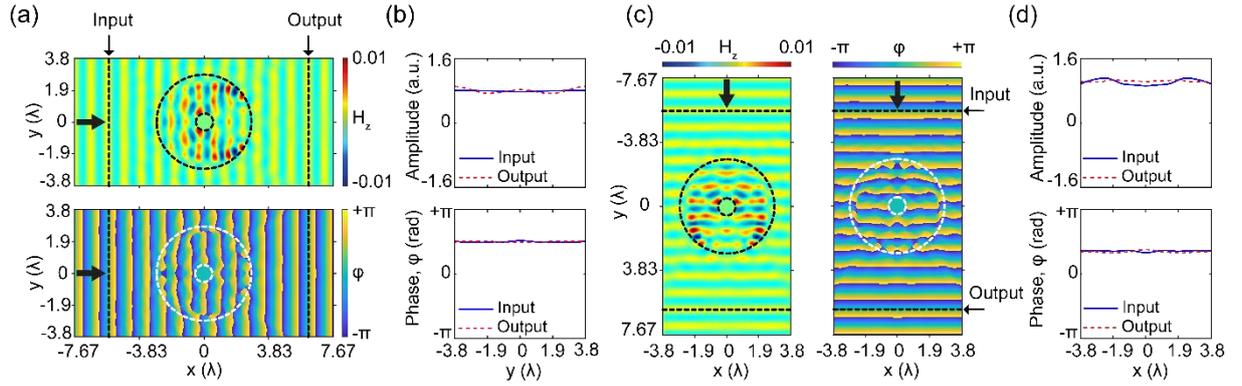

**Figure 3**. *The calculated (a) magnetic field and phase distributions, (b) cross-sectional amplitude and phase profiles at the front and back cross sections for x-direction injection for optimized structure. The calculated (c) magnetic field and phase distributions, (d) cross-sectional amplitude and phase profiles at the front and back cross sections for y-direction injection for optimized structure. The black arrows indicate the incident waves. The dashed circles represent the boundaries of the PEC material and the optimized structures. The position profiles at the input and output of the cross sections are signified by the vertical dashed lines. All calculations were performed at 10 GHz.*

The magnetic and phase field distributions and their cross-sectional profiles of the optimized bi-directional cloak are proposed in Figs. 3(a) and 3(b), respectively. As can be seen from Figs. 3(a) and 3(c), the optimized cloaking structure that covers the PEC object suppresses the field distortions/variations and successfully reproduces the incident plane wave at the back plane as well as by diminishing undesirable back reflections. From the wave propagation characteristic through the cloaking region, we can observe that the cloaking structure is designed in such a way that undesired reflections from the PEC are reduced into negligibly small values and corrects the distorted field at the output plane. In other words, the cloaking region operates as a transparent/anti-reflective coating effect that transmits the light through the PEC object without affecting its initial state. Furthermore, one can observe that propagating field is enhanced by guiding and confining inside of the cloak which results in phase matching behavior by effectively reducing the scattering field and restoring the wave fronts before and after the structure. As a result, substantial scattering cancellations are achieved with the optimal structure compared to the structures presented in Fig. 2.

As it was noted, the designed structure has 4-fold rotational symmetry as seen in Fig. 1. Here the rotational and mirror symmetry concepts are realized to provide quasi-omnidirectional invisibility behavior of the structure [23, 31, 33]. In this regard, the proposed structure is optimized considering these symmetry constraints and corresponding results are given in Figs. 3(c) and 3(d). As a result, one can observe the same bi-directional cloaking effect in the transverse y-direction with negligible field and phase profile distortions. Moreover, in both *x*- and *y*- incident directions, the field and phase profiles are almost identical to each other which justifies its bi-directional operation.

It is necessary to properly suppress the scattered field of the superimposed multiple modes to achieve invisibility cloak [49-51]. As can be observed from the results given in Figs. 3(a)-3(d), optimized cloaking structure suppresses higher order modes that enhance the scattering effect by reducing the observability of the PEC at operating frequency of 10 GHz. In this regard, it is informative to show scattering effect analysis of the optimized structure.

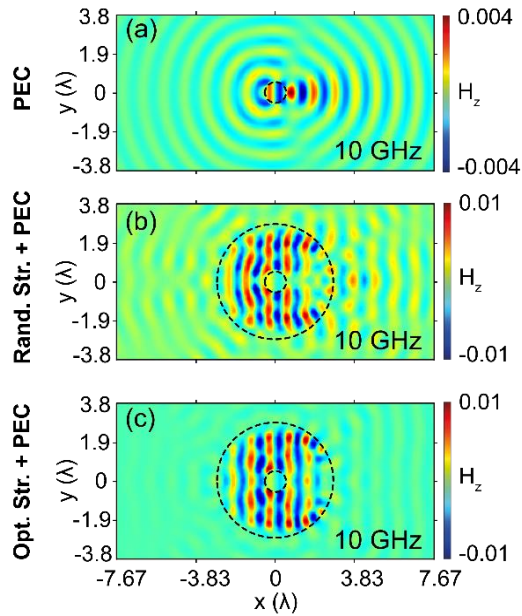

**Figure 4**. *The numerically calculated scattering field of (a) PEC, (b) randomly filled structure with PEC and (c) optimized structure with PEC at 10 GHz in the free space.*

In fact, the proposed cloaking structure presents scattering reducing effect in orthogonal (horizontal) and (vertical) directions as can be observed from the results in Fig. 3. It means that the optimization constraints and cost functions given in Eqs.1-3 are indirectly reduced backward, and forward scattering effect of the PEC coated by cloaking structure. To analyze cloaking performance quantitively, proposed design is evaluated by calculating the scattering fields of PEC with/without the proposed cloak and randomly generated structure. The corresponding results are presented in Fig. 4. The scattering effects of bare PEC, PEC coated with randomly generated and optimized cloak structures are gathered in Fig. 4. Here, Fig. 4(a) demonstrates the scattered field of the PEC. As expected, the perfect conductor generates a strong back-reflections and a shadow region in the forward direction which can also be inferred from Fig. 2(a). The effect of randomly distributed dielectric material (unit-cell) on incident light scattering is presented in Fig. 4 (b). Here, due to its randomness and inattentive distribution, we can observe strong diffraction effect which leads to undesired distortion of the incident wave. On the other hand, the strong scattering effect of a PEC is significantly reduced when it is coated by optimized cloak structure as seen in Fig. 4(c). The reduction occurs at both backward and forward scattering directions. The negligible appearance of the scattering field is covered by a wisely arrangement of spectral distribution of the dielectric material qualitatively demonstrates its invisibility performance.

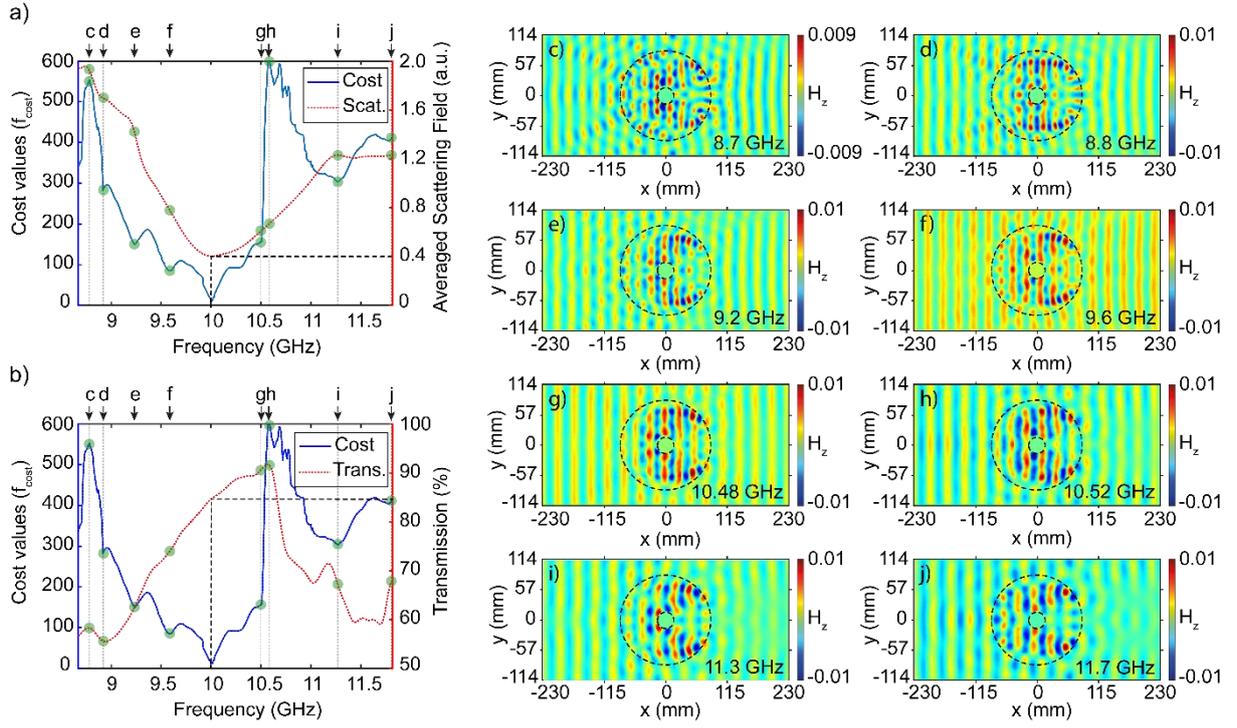

**Figure 5**. *The plots of cost values with (a) averaged scattered field values and with (b) transmission efficiency values for selected frequency intervals. The numerically calculated magnetic field distributions at (c) 8.7 GHz, (d) 8.8 GHz, (e) 9.2 GHz, (f) 9.6 GHz, (g) 10.48 GHz, (h) 10.52 GHz, (i) 11.3 GHz, and (j) 11.7 GHz.*

As stated previously, the transmissivity of the cloak is also crucial for it to operate efficiently. High transmittance of undistorted wave behavior is achieved during the optimization process because the cost function of the optimization is targeted to maximize transmission efficiency while reducing the distortions. Also, we should remind that the proposed structure is designed and optimized at a fixed design frequency of 10 GHz. It would be interesting to investigate the frequency response of the proposed structure in the vicinity of the design frequency. In addition, analyzing the variation of defined cost function and the performance characteristics of the structure such as scattering and transmission over operating frequencies of 8.8 GHz and 11.6 GHz can be instructive and beneficial to understand underlying mechanism of cloaking. For this reason, we prepared Fig. 5 where the trade-off between cost function, scattering, and transmission are analyzed. Here, the cost values ($f_{cost}$) for designed structure at different operating frequencies between 8.8 GHz and 11.6 GHz are calculated according to Eq. 1. Also, spatially averaged scattered fields [52] are calculated by averaging the fields for same operating frequency regime according to the following equation:

$$\overline{H_{scat}}(f) = \frac{\sum_y \sum_x (|H_{z,in}(x,y,f) - H_{z,free}(x,y,f)| + |H_{z,out}(x,y,f) - H_{z,free}(x,y,f)|)}{xy}, \quad (4)$$

where, $f$ represents interested frequency points in the given band, $H_{z,in}(x,y,f)$ and $H_{z,out}(x,y,f)$ indicate the values of magnetic field at *x,y* locations before and after cloak structure, respectively. Moreover, magnetic field values of the free space are denoted by

$H_{z,free}(x, y, f)$. In Fig. 5(a) we present the calculated cost and averaged scattering field values. Although the averaged scattered field was not an input parameter for the cost function, its values in the monitored frequency region demonstrate a similar trend. As expected, at the design frequency 10 GHz, the averaged scattered field takes the minimum value as the cost function. In Fig. 5(b) we superimposed cost values versus transmission efficiency of the structure for interested frequency band region. Here, the relation of the transmission efficiency and the cost values show logical dependency since increasing the transmission results in decreasing in the cost value as expected according to Eq. 1. Of course, both graphs do not present perfect harmony due to the different factors such as field and phase distortion that also affect the overall performance of the optimized structure. In this stage, it makes sense to make a detailed frequency response analysis of the proposed structure by considering specially selected frequency points. Here, we select 8 different frequency points (defined as *c, d, e, f, g, h, i, j*) which correspond to the local extremum values of the cost function as can be seen in Figs. 5(a) and 5(b). Corresponding frequency values of selected points are given as an inset in Figs. 5(c)-5(j). Semitransparent lines from *c* to *j* and green circle markers are defined in the figure plots 5(a) and 5(b) to show values of the averaged scattering field and transmission efficiencies that correspond to those extremum values of the cost function. The magnetic field distributions at those 8 frequency points are presented in Figs. 5(c)-5(j). As can be seen from Fig. 5(c), proposed structure shows strong distortion in back and forward fields for the operating frequency of 8.7 GHz. While approaching the design frequency of 10 GHz we see sequential improvement in suppression of backward and forward scatterings resulting in reducing distortion of wavefronts as demonstrated in Figs. 5(d)-5(f). At the same time, correction of distortions and increase in transmission efficiency cause a decrease in cost function. At the design frequency, cost function and average scattering field values are reaching their minimum value while transmission efficiency is reached 84%. On the other hand, increasing the frequency makes back-reflections apparent by causing strong distortions of the field as can be seen from Figs. 5(g)-5(j). Moreover, due to strong light matter interaction at the high frequencies cloaking structure demonstrates a strong forward shadowing effect by losing its phase matching ability which is explicitly seen from field distributions in Figs. 5(i) and 5(j). Although, we expect inverse relation between transmission and cost function, transmission efficiency shows increasing persistence until the frequency point of 10.52 GHz despite increasing cost function. This effect is caused by the defined cost function parameters where correction in field and phase distortion is dominated over the transmission effect during the optimization while balancing the trade-off between them. Even though the proposed structure is optimized at a fixed frequency, the results gathered in Fig. 5 shows that the optical cloaking is achievable with acceptable performance between frequencies of 9.5 GHz and 10.5 GHz. Due to the absence of resonance effect in the proposed approach, the multi-dimensional cloaking successfully operates over a certain bandwidth interval.

## 3. Experimental verification in microwave regime

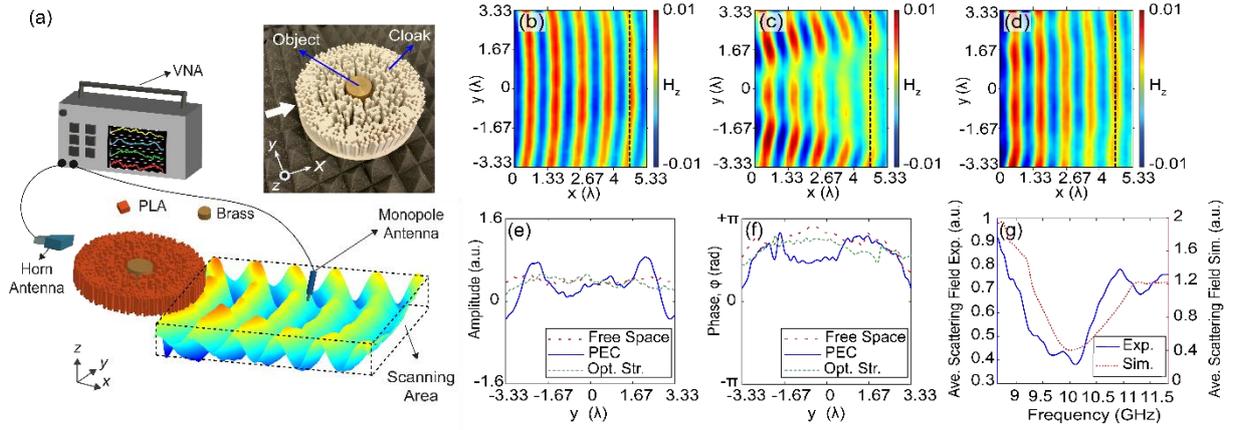

**Figure 6**. *(a) The schematic representation of the experimental setup with a photo of the fabricated cloaking structure and a brass object that is used as PEC. Experimentally measured magnetic field representations of (b) free space, (c) PEC and (d) optimized structure with PEC, respectively. (e) Experimentally measured cross sectional amplitude profile and (f) cross sectional phase profile. (g) Averaged scattering field values for selected frequency intervals. The black vertical dashed lines are signified the positions of the cross sections.*

Experimental verification of the cloaking performance of the proposed structure is realized at microwave frequency regime. The optimized cloak is fabricated by a 3D printing technique which utilizes (polylactic acid) PLA material which is a widely used plastic filament material for rapid manufacturing. The PLA material provides a permittivity value of PLA $\varepsilon_{PLA} = 2.4$ at microwave frequencies between 8 GHz –12 GHz, according to the Nicolson–Ross and Weir measurement method [53]. In place of the PEC object that is intended to be cloaked, a cylindrical object made of brass material with a diameter of 34.6 mm and a height of 45 mm is used in the experiment. Brass is the compound of copper and zinc materials, which has scattering properties at microwave frequencies between 8 GHz –12 GHz. Throughout the experimental process, the Agilent E5071C ENA vector network analyzer is used to generate and measure electromagnetic waves. The generated microwaves at an operating frequency of 10 GHz were directed towards the cloaking structure by the horn antenna in front of the structure as seen from the experimental setup with an artistic illustration of propagating wave behind the cloak in Fig. 6(a). Also, the monopole antenna is connected to the same network analyzer and located on a motorized stage to measure the magnetic field distribution behind the designed cloaking structure. It should be noted that an aperture antenna is placed at a distance to provide a planar wavefront of the incident wave at the front surface of cloaking structure because horn antenna generates a Gaussian profiled wave with spherical wavefronts. The photographic illustration brass cylindrical object coated by fabricated cloaking structure is shown as an inset in Fig. 6(a). To evaluate the cloaking performance of the structure qualitatively and quantitatively, three different experimental cases are considered. Firstly, free space propagation of the incident wave is measured in a defined area of measurement (scanning area) located behind the position of the cloak. The corresponding magnetic field distribution of the free space propagation at the design 10 GHz frequency is presented in Fig. 6(b). Later, we positioned the brass object only and measured the magnetic field. Finally, we placed the designed cloak to surround the brass object and realized same measurements in the same scanning area at the frequency of 10 GHz. The corresponding magnetic fields are presented in Figs. 6(c) and 6(d), respectively. As can be seen from Fig. 6(c), a cylindrical brass object strongly scatters the incident wave and the resulting wavefront behind the structure is divided into two branches (the appearance of the shadow is explicitly visible). On the other hand, the cloaking structure pull the scattered fields together by conveying them around the brass object to reproduce a wave resembling the plane wave. In Figs. 6(e) and 6(f), the cross-sectional amplitude and phase profiles at the output are plotted for the cases of free space, PEC only and PEC coated by cloak, respectively. Here, one can see that cylindrical brass strongly scatters the incident plane wave and leads to strong variations in both the magnetic field and phase distributions. On the other hand, proposed cloaking structure minimizes/suppresses the field variations and successfully recovers the incident plane wave behind the structure. To present scattering suppressing performance of the cloak, the average scattered field is calculated by Eq. 4 using the data from the experiments. Measured and numerically calculated average scattering field variations with respect to the microwave frequencies are superimposed in Fig. 6(g). Here, we can see that minimum scattering value appeared at around 10 GHz which is parallel to the numerical results. From the experimental results, it can be concluded that optimized cloaking structure exhibit an invisibility effect for highly scattering objects by recovering the distorted field at the observation area.

## 4. Further discussion: the concept of multi-directional cloaking

Previously in this study, we introduced the design and optimization of bi-directional cloaking structure by using rectangular shaped unit cells. Generally, the numerical design and manufacturing of rectangular shape unit cells with defined structural parameters are comparably simple and cost effective [33]. It is crucial to note that rectangular shaped unit cells are suitable only for one and two directional cloaking structure designs because of perfect radial alignment between mirror symmetrized slices. However, the number of cloaking directions greater than two (internal angle of the mirror symmetrized slices smaller than 45°) is not physically/geometrically achievable with rectangular shaped unit cells. For this reason, to realize multi-directional cloaking we further propose exploiting of unit cells having shape of annular sectors instead of rectangular ones. The schematic representation of the optimized 3-directional cloaking structure composed of annular sectors is presented in Fig. 7(a). Here, the number of mirror-symmetrized slices is increased to 12 according to the number of injection directions. The incident wave injection angles are defined as 0°, 60° and 120° degrees as can be seen in Fig. 7(a). The overall size of the cloaking structure is kept the same as the bi-directional cloaking structure proposed in Fig. 1(a). A circle is divided into 12 slices and one of the slices is introduced to the algorithm to define the existence of unit cells that construct the slice. After determining the existence of unit cells thanks to GA integrated with FDTD in the slice, again mirror symmetry approach was used for the optimized slice as a design approach. Then, it is rotated to construct the optimal circular cloaking structure. To construct the optimized slice, we determine the radius of the circle (amount of circle) based on the construction constraints (3D printer) and degree of the slice to construct the intended angle. The degree of each unit cell is arranged such that when one ends the other starts, and the difference of both radii is determined to correspond to $0.1\lambda$.

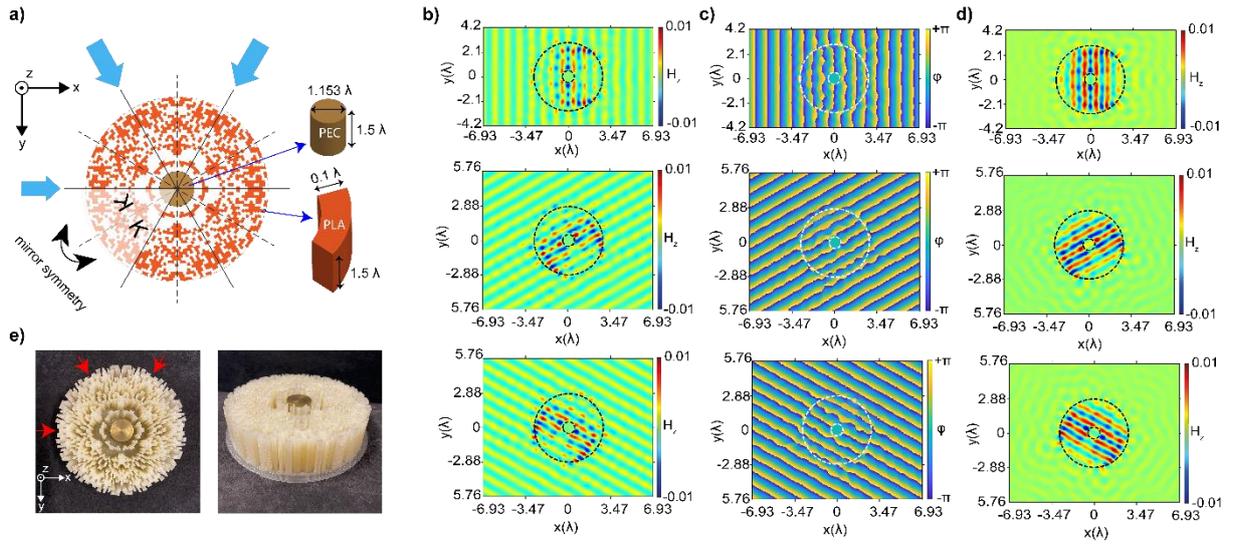

**Figure 7**. (a) Schematic representation and the design approach of the cloaking structure that operates in three directions. Blue arrows demonstrate the incident wave directions with corresponding incident angles. (b) Calculated magnetic field, (c) phase distributions and (d) scattering field distributions for incident directions of 0°, 60°, and 120°. Black and white dashed lines demonstrate the boundaries of the PEC and cloaking structure. (e) Photographic illustration of the 3D printed cloaking structure and a brass object inside at top and perspective views. Again, red colored arrows demonstrate the incident wave directions.

As can be seen from Figs. 7(b) and 7(b), the optimized cloaking structure that covers the PEC object suppresses the field and phase distortions/variations and successfully reproduces the incident plane wave at the back plane as well as by diminishing undesirable back reflections. From the wave propagation characteristic through the cloaking region, we can observe that the cloaking structure is designed in such a way that undesired reflections from the PEC are reduced into negligibly small values and corrects the distorted field at the output plane. In other words, the cloaking region operates as a transparent/anti-reflective coating effect that transmits the light through the PEC object without affecting its initial state. Furthermore, one can observe that propagating field is enhanced by guiding and confining inside of the cloak which results in phase matching behavior by effectively reducing the scattering field and restoring the wave fronts before and after the structure. From Fig. 7(c), it can be said that strong scattering cancellations are achieved with the optimal structure for three different directions of incident. It should be noted that by appropriately the dimensions of the design and selecting appropriate material, the same approach can be applied to different nanomaterials. The structure can be realized via different manufacturing techniques such as e-beam lithography or direct laser writing.

## 5. Conclusion

In the presented study, we show the design and experimental proof of an optical cloaking structure for multi-directional hiding of a perfectly electric conductor (PEC) object from incident light wave operating at the design frequency which is 10 GHz. The structure has a circular shape which is discretized into square elementary cells that can be either non-magnetic/all dielectric polylactide (PLA) material or air which is meta-heuristically determined by the Darwinian concept of natural selection method. In other words, the dielectric modulation around the highly reflective scattering PEC object is determined by an optimization process for multi-directional cloaking purposes. Moreover, to obtain the multi-directional effect of the cloaking structure, an optimized slice is mirror symmetrized through a radial perimeter. Three-dimensional (3D) finite-difference time-domain method is integrated with genetic optimization to achieve cloaking design. The main objective of the designed cloaking structure is to suppress the reflected light and reproduce the transmitted light into the plane wave by reducing the amplitude fluctuations in the cross-sectional field and phase profiles. In addition, optical power transmission characteristic of the directional cloaking structure is considered during the concealing of PEC object by cloaking structure. To quantify cloaking effect spatially averaged scattered fields are also calculated. The proposed structure is fabricated by 3D printing of PLA material and the brass metallic alloy is used as a perfect electrical conductor for microwave experiments. A good agreement between numerical and experimental results is achieved. Using nanotechnology, our scale proof-of-concept research will take the next step towards the creation of "optical cloaking" devices in nanoscale.